\documentclass[sigconf]{acmart}
\AtBeginDocument{%
  }

\copyrightyear{2025}
\acmYear{2025}
\setcopyright{rightsretained}
\acmConference[RecSys '25]{Proceedings of the Nineteenth ACM Conference on Recommender Systems}{September 22--26, 2025}{Prague, Czech Republic}
\acmBooktitle{Proceedings of the Nineteenth ACM Conference on Recommender Systems (RecSys '25), September 22--26, 2025, Prague, Czech Republic}\acmDOI{10.1145/3705328.3748143}
\acmISBN{979-8-4007-1364-4/2025/09}

\begin{document}

\title{Scaling Generative Recommendations with Context Parallelism on Hierarchical Sequential Transducers}

\author{Yue Dong}
\email{yoyoyod@meta.com}
\affiliation{%
  \institution{Meta Platforms}
  \city{Menlo Park}
  \state{California}
  \country{USA}
}

\author{Han Li}
\email{hanli@meta.com}
\affiliation{%
  \institution{Meta Platforms}
  \city{Menlo Park}
  \state{California}
  \country{USA}
}

\author{Shen Li}
\email{shenli@meta.com}
\affiliation{%
  \institution{Meta Platforms}
  \city{Menlo Park}
  \state{California}
  \country{USA}
}

\author{Nikhil Patel}
\email{nikhilp@meta.com}
\affiliation{%
  \institution{Meta Platforms}
  \city{Menlo Park}
  \state{California}
  \country{USA}
}

\author{Xing Liu}
\email{xingl@meta.com}
\affiliation{%
  \institution{Meta Platforms}
  \city{Menlo Park}
  \state{California}
  \country{USA}
}

\author{Xiaodong Wang}
\email{xdwang@meta.com}
\affiliation{%
  \institution{Meta Platforms}
  \city{Menlo Park}
  \state{California}
  \country{USA}
}

\author{Chuanhao Zhuge}
\email{czhuge@meta.com}
\affiliation{%
  \institution{Meta Platforms}
  \city{Menlo Park}
  \state{California}
  \country{USA}
}

\renewcommand{\shortauthors}{Dong et al.}

\begin{abstract}
Large-scale recommendation systems are pivotal to process an immense volume of daily user interactions, requiring the effective modeling of high cardinality and heterogeneous features to ensure accurate predictions. In prior work, we introduced Hierarchical Sequential Transducers (HSTU), an attention-based  architecture for modeling high cardinality, non-stationary streaming recommendation data, providing good scaling law in the generative recommender framework (GR). Recent studies and experiments demonstrate that attending to longer user history sequences yields significant metric improvements. However, scaling sequence length is activation-heavy, necessitating parallelism solutions to effectively shard activation memory. 
In transformer-based LLMs, context parallelism (CP) is a commonly used technique that distributes computation along the sequence-length dimension across multiple GPUs, effectively reducing memory usage from attention activations. In contrast, production ranking models typically utilize jagged input tensors to represent user interaction features, introducing unique CP implementation challenges.  In this work, we introduce context parallelism with jagged tensor support for HSTU attention, establishing foundational capabilities for scaling up sequence dimensions. Our approach enables a 5.3× increase in supported user interaction sequence length, while achieving a 1.55× scaling factor when combined with Distributed Data Parallelism (DDP).

\end{abstract}


\maketitle


\section{Introduction}
Modern attention-based architectures enable substantially larger-scale processing, thereby making it possible to leverage significantly more data within recommendation systems. By learning from considerably larger volumes of user impressions and interactions across various product surfaces, these models achieve deeper personalization and more accurate recommendations than previously possible \cite{chang2023twin, pancha2022pinnerformer, sun2019bert4rec}. To effectively learn from longer user interaction history sequences, previous research introduced a novel attention-based encoder design, known as hierarchical sequential transducers (HSTU) \cite{zhai2024actionsspeaklouderwords}. To efficiently process the variable-length user sequence input commonly seen in recommender sequence learning, the encoder adopts a jagged tensor implementation optimized with specialized Triton kernels.

Using this effective architecture, we identify a clear path to scale the model with model quality measured by normalized entropy (NE) improvements. Scaling the model includes learning from longer sequence length; however, this scaling is constrained by substantial activation memory requirements. To overcome this limitation, we explore context-parallelism (CP) \cite{liu2023ringattention}: a parallel computing technique that partitions long sequences across multiple GPUs, reducing per-device activation memory requirements, and thereby enabling greater scalability. Nonetheless, context-parallelism methods designed for standard transformer-based large language models (LLMs) cannot be directly applied due to fundamental differences in how jagged tensor inputs are represented and processed within the HSTU framework. In this paper, we introduce jagged tensor context-parallelism, an adaptation specifically designed for the HSTU architecture, serving as our primary technical contribution to efficiently scale sequence lengths and enable further improvements in recommendation system performance.

\section{Methodology}
In this section, we present the main components of the generative recommender encoder, including the HSTU architecture and the jagged tensor representation. We outline how these elements differ from traditional transformer-based LLM architectures, and discuss our approach to efficiently implementing context-parallelism tailored specifically to the unique characteristics of our system.

\subsection{Sequential Transducer Attention}
The formulation of the HSTU encoder is defined by the following equation:

\[
\text{HSTUAttention}(Q,K,V) = \text{Mask}\left(\text{SiLu}\left(\frac{QK^T + \text{Bias}}{\sqrt{d_k}}\right)\right)V
\]

We highlight that a critical component is the customized timestamp bias, which is generated based on the timestamps with learnable weights. It first computes the differences between consecutive elements in \textit{ts}, then performs the transformation followed by the grouping. These bucket indices are then used to select weights from \textit{ts\_weights}, which is a learnable parameter tensor initialized with values drawn from a normal distribution.

\subsection{Jagged Tensors}
Unlike dense features, which typically have fixed sizes, sparse features frequently observed in training examples exhibit high sparsity and variable sequence lengths. To efficiently represent these sparse features, we preprocess them into jagged tensors containing both feature \textit{values} and metadata, such as \textit{offsets} and \textit{max\_lengths}. These jagged tensors are subsequently provided as inputs to the attention operations. We use the TorchRec \cite{torchrec2024} implementation of the jagged tensor.

\subsection{Context-Parallelism Implementation}
The key idea of Context Parallelism (CP) is to leverage the block-wise computation of self-attention to distribute long sequences across multiple devices to reduce memory usage. A typical workflow includes sharding Q, K, V on sequence length dimension so each device only holds 1/N of the data; each device then independently performs local attention computations before exchanging intermediate results in a block-wise manner to obtain the final attention outputs. To achieve efficient processing, ring-based gathering is utilized, with the intention to fully overlap communication with computation.

\begin{figure}[htbp]
    \centering
    \includegraphics[width=\linewidth]{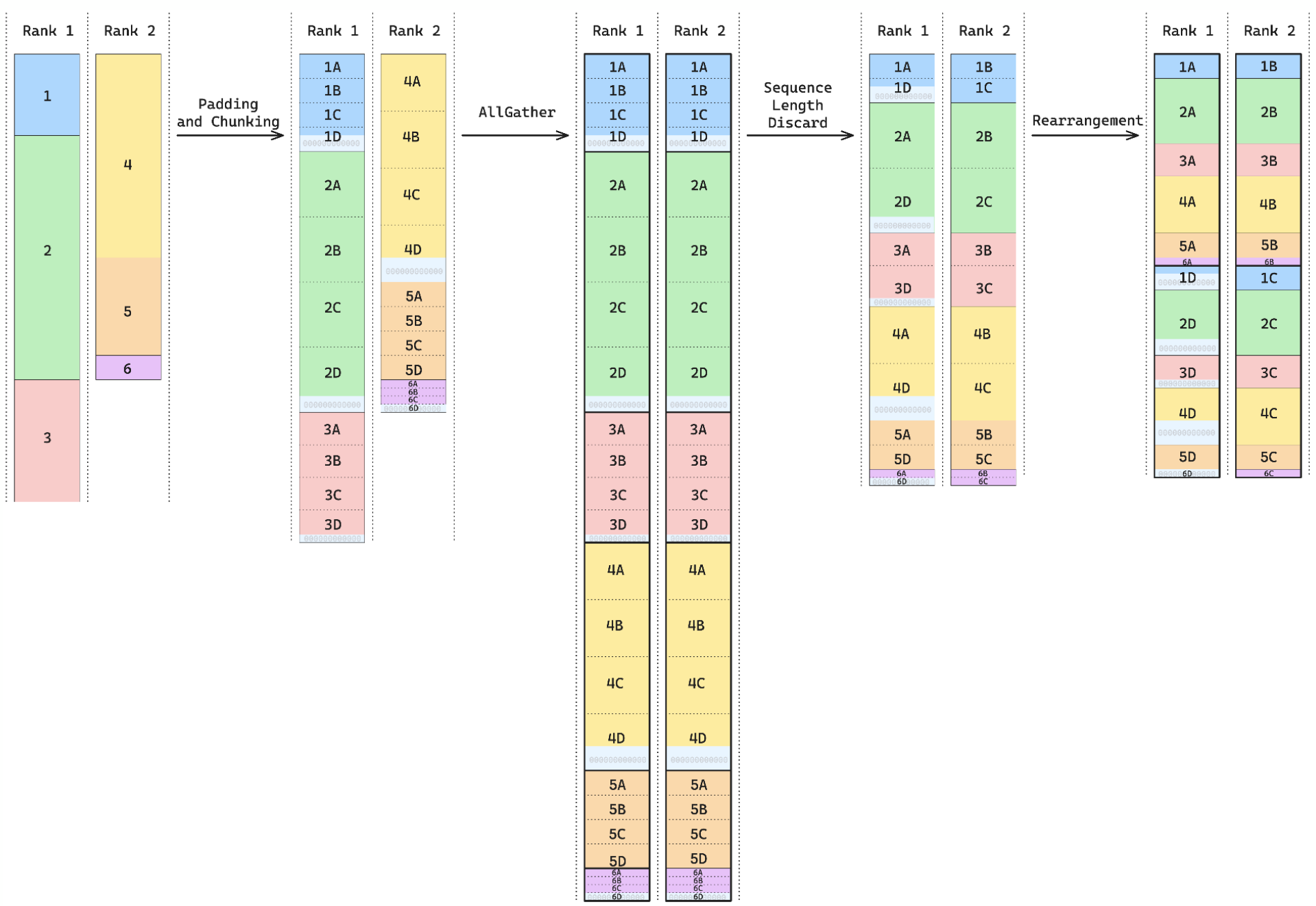}
    \caption{Jagged Tensor AllGather Process}
    \label{fig:jagged-allgather}
\end{figure}

In our implementation, we focus on scenarios where CP is introduced as an additional dimension of parallelism alongside DDP. A standard pre-processing interation with DDP is to \textit{AllGather} along the batch dimension, followed by an even split along the sequence dimension. However, the inherently dynamic nature of jagged tensors — characterized by highly variable sequence lengths, common in recommendation system training — introduces significant challenges in efficiently implementing the communication required for this \textit{AllGather} operation. As can be seen in Figure~\ref{fig:jagged-allgather}, both the communication volume and peak memory usage at the intermediate stage are dominated by the total concatenated sequence length. 

\begin{figure}[htbp]
    \centering
    \includegraphics[width=\linewidth]{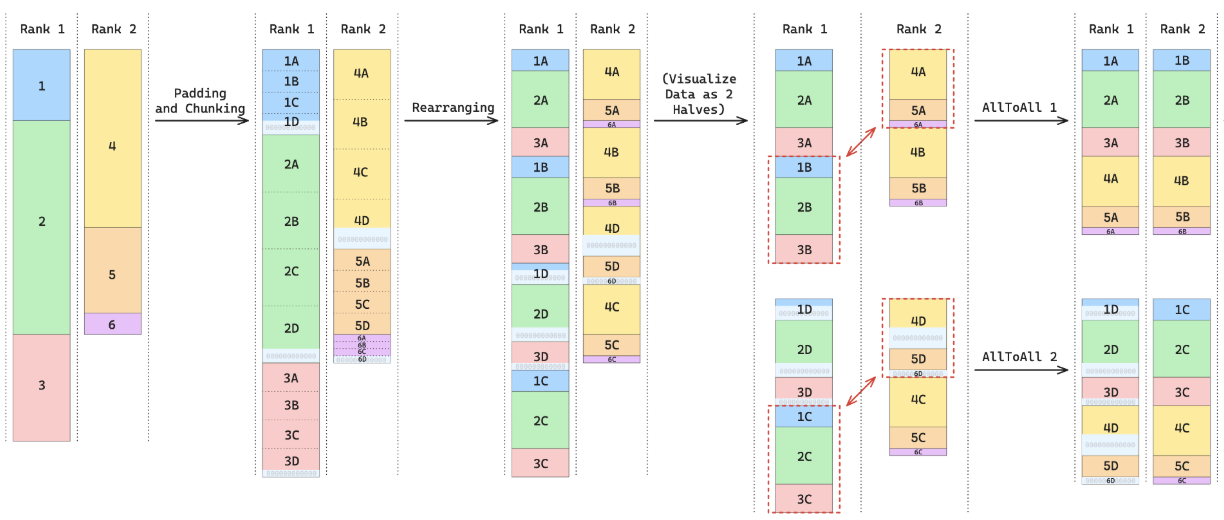}
    \caption{Jagged Tensor AllToAll Process}
    \label{fig:jagged-a2a}
\end{figure}

In order to improve communication and memory efficiency, we replaced the \textit{AllGather} with \textit{AllToAll} to directly send relevant chunks from each sample to each rank, thereby saving memory by avoiding full copy of data on each rank. This is illustrated in Figure~\ref{fig:jagged-a2a}. With this technique, we are able to significantly reduce the peak memory compared to the \textit{AllGather} method by >60\% and improve the training throughput measured by examples per-second by \textbf{>2x}.

\begin{figure}[htbp]
    \centering
    \includegraphics[width=\linewidth]{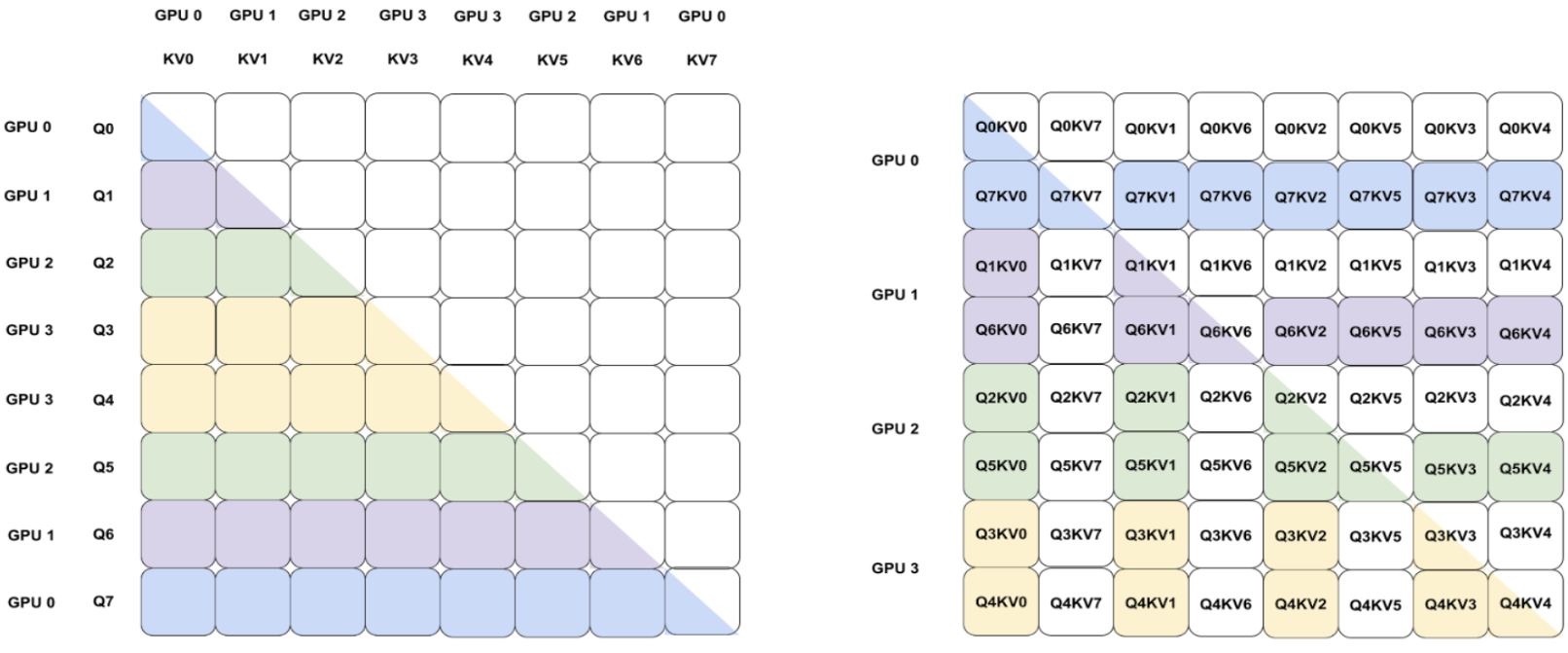}
    \caption{Load Balanced CP with 2*CP\_size Mini-chunks}
    \label{fig:lb}
\end{figure}

Another common challenge encountered in CP is load balancing. In both standard transformer-based LLM attention and HSTU attention, a mask is typically applied to restrict which positions in the input sequence can attend to each other during self-attention, thus preventing information leakage from future positions. The most frequently used mask is the “causal” (or “lower-triangular”) mask. However, using a triangular mask inherently leads to computational imbalance across different GPU ranks. Several strategies~\cite{li2024distflashattn, transformerengine2025} have been proposed to mitigate this load imbalance. We adopt an approach similar to that used in \texttt{TransformerEngine} \cite{transformerengine2025}. For each chunk, we further partition into 2 mini-CP chunks along sequence-length dimension, and assign chunks \textit{(i, 2xCP - 1 - i)} to \textit{rank i}. This ensures balanced computational loads across GPU ranks, as illustrated in Figure~\ref{fig:lb}. To effectively implement this approach, we employ efficient data shuffling with custom Triton kernels designed for memory reordering, thereby avoiding significant overhead from chunking and concatenation operations. This technique further enhances overall training throughput by \textbf{37\%}.

\section{Experiments and Conclusions}

We conducted experiments on Jagged CP with HSTU, running on NVidia H100 GPUs with 80GB HBM. The baseline model was executed using DDP without activation checkpointing, achieving a maximum sequence length of 3K (utilizing 89\% of available memory). Scaling beyond this point resulted in out-of-memory (OOM) errors. With the adoption of CP, we successfully scaled to significantly longer sequence lengths proportional to the CP size, as illustrated in Table~\ref{tab:max-seq-vs-cp}. 

\begin{table}[ht]
\centering
\begin{tabular}{lc}
\toprule
\textbf{CP Size}   & \textbf{Maximum Sequence Length} \\
\midrule
Without CP         & 3072  \\
CP=2               & 4096  \\
CP=4               & 7168  \\
CP=8               & 16384 \\
\bottomrule
\end{tabular}
\caption{Maximum sequence length supported vs.\ CP size}
\label{tab:max-seq-vs-cp}
\end{table}

We also evaluate the scaling efficiency by measuring the achieved scaling factor. Ideally, the scaling factor should increase linearly with the number of GPUs. In our experiments, we observed scaling efficiency ranging between 1.33× and 1.55×.

\begin{table}[ht]
\centering
\begin{tabular}{lcc}
\toprule
\textbf{Case} & \textbf{Scaling Factor}\\
\midrule
Ideal case                       & 2.00×            \\
DDP only                         & 1.6–1.7×         \\
DDP + CP (cp\_size=2)            & ~1.33×            \\
DDP + CP (cp\_size=2, 2× batch)  & ~1.55×            \\
\bottomrule
\end{tabular}
\caption{Scaling factors and raw QPS under different parallelism setups (baseline: 24 K QPS).}
\label{tab:scaling-qps}
\end{table}

We break down how each performance optimization technique described above contributes to improving training throughput, measured in queries (examples) per second (QPS). First, replacing \textit{AllGather} with \textit{AllToAll} eliminated redundant memory usage and communication overhead, resulting in a 2.7× increase in QPS. Next, the custom Triton kernels for memory reordering—an essential step for load balancing—which avoided costly chunking and concatenation, provides additional 37\% improvement. Finally, we addressed GPU/CPU synchronization overhead introduced by jagged tensor communication. Specifically, to compute sequence lengths for collective communication, we previously relied on \textit{.item()} calls that triggered device-host sync. By asynchronously copying the offset tensor to the CPU and delaying synchronization until absolutely necessary, we achieved a further ~2\% QPS gain.

\begin{figure}[htbp]
    \centering
    \includegraphics[width=\linewidth]{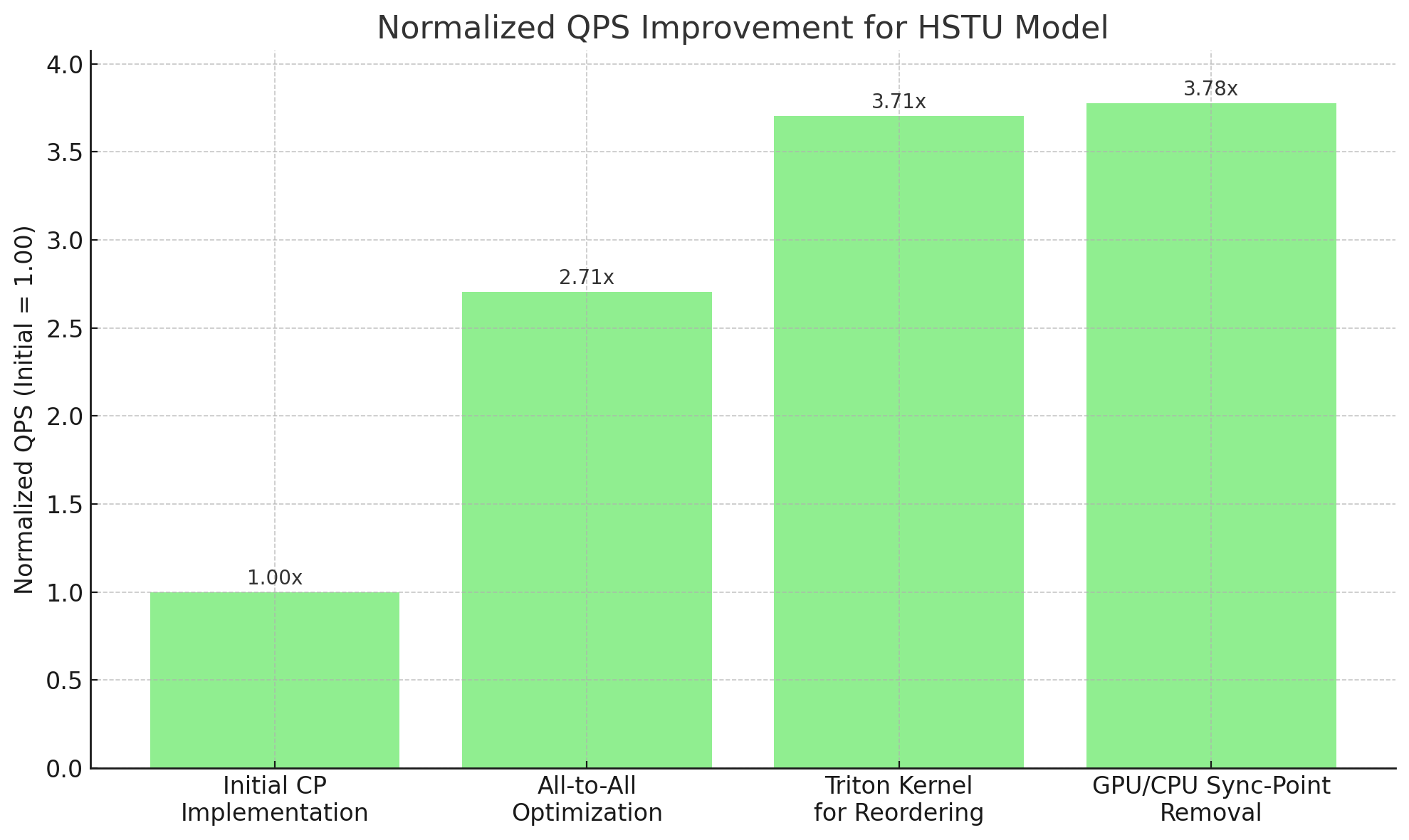}
    \caption{Normalized QPS Gains by Optimization Techniques}
    \label{fig:qps-improve}
\end{figure}

We have also identified additional opportunities to further enhance scaling efficiency. Currently, the performance bottleneck lies in the attention kernel, which was optimized for shorter sequences. A promising opportunity involves fusing these segmented kernels into larger, consolidated kernels. Our preliminary headroom study indicates that implementing this fusion can increase the scaling factor at the same batch size from 1.33× to approximately 1.5×–1.6×. 

In summary, our work on jagged tensor context-parallelism (CP) for HSTU enables efficient scaling of generative recommender models to process significantly longer feature sequences, thereby enhancing the overall quality and effectiveness of recommendation systems.

\begin{acks}
This work reflects the collaborative efforts of many contributors. We are especially grateful - in alphabetical order by first name - to Daniel Johnson, David Berard, Jade Nie, Linjian Ma, Weiwei Chu, Xinfeng Xie, Yuchen Hao, and others whose insightful feedback and technical input were instrumental throughout the project. We would also like to thank Adnan Aziz, Ajit Mathews, Bi Xue, Daisy Shi He, Emad El-Haraty, Jiaqi Zhai, and Min Ni for their leadership and support.
\end{acks}


\bibliographystyle{ACM-Reference-Format}



\appendix

\end{document}